\documentclass[aps]{revtex4}
\newcommand{\bb}{\begin{eqnarray}}
\newcommand{\ee}{\end{eqnarray}}
\begin{document}
\title{{Hawking temperature and higher order calculations}}
\author{Bhramar Chatterjee}\email{bhramar.chatterjee@saha.ac.in}
\author{P. Mitra}\email{parthasarathi.mitra@saha.ac.in}
\affiliation{Saha Institute of Nuclear Physics\\
Block AF, Bidhannagar\\
Calcutta 700 064}
\begin{abstract}
Hawking radiation has recently been explained by using
solutions of wave equations across black hole horizons 
in a WKB approximation. Higher order
calculations using both usual and non-singular coordinates
are found to change the solution for zero spin, but this change
is not an alteration of the Hawking temperature. For spin 1/2,
there is no correction to the simplest form of the solution.
\end{abstract}
\maketitle
\bigskip

A classical black hole has a  horizon  beyond which  nothing
can escape, but quantum theory alters the situation as shown by
the theoretical discovery of radiation from black holes \cite{Hawk}. 
For a Schwarzschild black hole, the radiation, which is thermal, has a
temperature
\bb
T_H={\hbar\over 4\pi r_h}={\hbar\over 8\pi M},
\ee
where $r_h$ gives the location of the horizon in standard coordinates
and $M$ is the mass of the black hole.
This was derived by considering quantum massless particles in a
Schwarzschild background geometry. The derivation being quite
complicated, attempts have been made to understand the process of
radiation by other methods. In \cite{HH}, a path integral study
was made, and analytic continuation in complex time used to 
provide further evidence
for the temperature $T_H$. Furthermore, the propagator in the
Schwarzschild background was shown \cite{HH} to have a periodicity
in the imaginary part of time with period $4\pi r_h= 8\pi M$,
again suggesting the same temperature. There is also an argument
involving a conical singularity on passing to imaginary time, which
can only be avoided if the standard Hawking temperature is chosen. 

Later, other attempts were made to understand the emission of particles
across the horizon as a quantum mechanical tunnelling process \cite{x,PW}.
A WKB treatment was used to reproduce the standard temperature 
$T_H$ \cite{zerbini}. Some ambiguities were resolved in \cite{mitra}
and clarified using non-singular coordinates in \cite{cgm}. 
Various related discussions can be found in \cite{tun}.
We shall discuss here corrections to the wave function
when higher orders in Planck's constant are considered.
 
A massless spin zero particle in the Schwarzschild background is described
by the Klein-Gordon equation
\bb
{\hbar^2\over\sqrt{-g}}\partial_\mu(g^{\mu\nu}\sqrt{-g}\partial_\nu\Phi)=0.
\ee
One writes
\bb
\Phi=\exp(-{i\over\hbar}S)
\ee
and obtains, to the leading order in $\hbar$, the equation
\bb
g^{\mu\nu}\partial_\mu S\partial_\nu S=0.
\ee
If we consider spherically symmetric solutions and
use separation of variables to write
\bb
S=Et+C+S_0(r),
\ee
where $C$ is a constant {\it arising from the integration of} 
${\partial S\over \partial t}=E$, 
the equation for $S_0$ becomes
\bb
-{E^2\over 1-\frac{r_h}{r}} + (1-\frac{r_h}{r})S_0'(r)^2=0
\ee
in the Schwarzschild metric. The formal solution of this equation is
\bb
S_0(r)=\pm E\int^r{dr\over  1-\frac{r_h}{r}}.
\ee
The two solutions correspond to the
fact that there are incoming and outgoing solutions. There is
a singularity at the horizon $r=r_h$, which is taken care of 
by skirting the pole and changing $r-r_h$ to  $r-r_h-i\epsilon$.
This yields
\bb
S_0(r)=\pm E[r+r_h\cdot i\pi+r_h\int^r dr P(\frac{1}{r-r_h})],
\ee
where $P()$ denotes the principal value. 
Let us write the incoming and outgoing solutions for $r>r_h$ as
\bb
S_{in}&=&Et+C+E[r+r_h\cdot i\pi+r_h\int^r dr P(\frac{1}{r-r_h})],\nonumber\\
S_{out}&=&Et+C-E[r+r_h\cdot i\pi+r_h\int^r dr P(\frac{1}{r-r_h})],
\ee
The real part of $C$ is arbitrary, but
if the imaginary part is ignored, strange consequences follow.
The imaginary part of $S_{out}$ implies a decay factor $\exp (-\pi r_hE/\hbar)$ 
in the amplitude, and a factor $\exp (-2\pi r_hE/\hbar)$ in the probability,
which is not consistent with the known Hawking temperature. Further, the
imaginary part of $S_{in}$ also implies a decay factor $\exp (-\pi r_hE/\hbar)$ 
in the amplitude, and a factor $\exp (-2\pi r_hE/\hbar)$ in the probability:
the motion is in the inward direction, so this factor in the probability is 
{\it not} $\exp (+2\pi r_hE/\hbar)$ (which is greater than unity)
as might appear at first sight. Thus absorption is
inhibited, as in a white hole \cite{cgm}. To sort this out, one 
should use coordinates which are smooth across the horizon
\cite{cgm}, but a simple method is to
make the imaginary part of $C$ cancel the imaginary part
of $S_{in}$, so as to ensure that the incoming probability
is unity {\it in the classical limit} -- when there is no reflection and everything 
is absorbed -- instead of zero or infinity \cite{mitra}. 
The need for the boundary condition of unit incoming probability
arises because the {\it Schwarzschild coordinates do not distinguish between
a black hole and a white hole}, and a specific choice has to be made
explicitly \cite{cgm}. Then
\bb
C&=&-i\pi r_hE + (Re~C),\nonumber\\
S_{out}&=&Et-E[r+r_h\cdot 2i\pi+r_h\int^r dr P(\frac{1}{r-r_h})]+ (Re~C),
\ee
implying a decay factor $\exp (-2\pi r_hE/\hbar)$ in the amplitude,
and a factor $\exp (-4\pi r_hE/\hbar)$ in the probability,
in conformity with the standard value of the Hawking temperature.

In the above calculation, $S$ has been treated only to the lowest order
in Planck's constant. We shall examine what happens in higher orders.
For the Schwarzschild metric, 
the full equation for a spherically symmetric $S$ reads
\bb
r^2[(1-\frac{r_h}{r})^{-1}\dot S^2-(1-\frac{r_h}{r})S'^2]=
-i\hbar[r^2(1-\frac{r_h}{r})^{-1}\ddot S-(r^2(1-\frac{r_h}{r})S')'].
\ee
It may be pointed out that \cite{imajh} does not include the $r^2$ factors in this 
equation which come from $\sqrt{-g}$ in three space dimensions
and thus differs from us. We write 
\bb
S=Et+C+S_0(r)+S_1(r)+S_2(r)+...,
\ee
where $S_n(r)$ is of $n^{th}$ degree in $\hbar$.
The terms of $n^{th}$ degree in $\hbar$ for $n>0$ yield
\bb
r^2(1-\frac{r_h}{r})\sum_{r=0}^nS_r'S_{n-r}'=-i\hbar[r^2(1-\frac{r_h}{r})S_{n-1}']'.
\ee
This equation expresses $S_n(r)'$ in terms of lower order $S_n'$-s
and can be solved recursively. In particular,
\bb
S_1'=-i\frac{\hbar}{r},
\ee
\bb
S_2'=\mp\frac{\hbar^2r_h}{2Er^3}.
\ee
Odd order $S_n(r)$ are seen to be purely imaginary and 
of the same sign for outgoing and incoming solutions. 
Even order $S_n(r)$ are real and differ in sign for the two
solutions.

We may also consider advanced Eddington-Finkelstein coordinates which are
non-singular at the horizon. In this case, the equation for 
a spherically symmetric solution reads
\bb
(r^2-rr_h)S'^2+2r^2ES'+i\hbar[2rE+(r^2-rr_h)S''+(2r-r_h)S']=0.
\ee
To zeroeth order, one gets two solutions for $S'$, corresponding to incoming 
and outgoing cases \cite{cgm}:
\bb
S'_0=0\quad {\rm or} \quad -{2rE\over r-r_h}.
\ee
The incoming solution shows full absorption and the outgoing solution
leads to the standard Hawking temperature without any need for a $C$.
The next terms are
\bb
S_1'=-i\frac{\hbar}{r},\quad\quad
S_2'=\mp\frac{\hbar^2r_h}{2Er^3}.
\ee
These are identical with the corrections found above for Schwarzschild
coordinates. In fact, all corrections are as in that case.

As regards the significance of these corrections to $S$, we first note that
$S_1$ is not really a quantum correction, because $S_1/\hbar$ is
independent of $\hbar$. 
\bb
S_1/\hbar=-i \ln r,
\ee
meaning that the wave function $\Phi$ gets a multiplicative factor $1/r$, 
which was only to be expected in three space dimensions. 
$S_2$, like other even order corrections, is real
and does not play any significant r\^{o}le. The first significant
correction is 
\bb 
S_3=i\frac{\hbar^3r_h}{4E^2r^3}(1-\frac{r_h}{r})
\ee
which alters the amplitude of $\Phi$ for both incoming and outgoing
solutions. However, this alteration depends not only on $E$ through $1/E^2$ 
but also on $r$ and vanishes at the horizon $r=r_h$ as well as at infinity. 
It cannot be regarded as an alteration of the Hawking temperature.
In general, 
\bb
S_n\propto E^{1-n},
\ee 
which changes the distribution in a complicated fashion.

It is interesting to see what happens if the Klein Gordon equation is
replaced by the Dirac equation. A spin connection has to be introduced.
In the massless case, two component wave functions are appropriate.
There are no solutions independent of the angular variables \cite{brill}, 
but one can take the angular part to be $1/\sqrt{\sin\theta}$. In this
simple case, the wave function can be written as
\bb
\Psi={(1-\frac{r_h}{r})^{-1/4}\over r \sqrt{\sin\theta}}e^{-iEt/\hbar}
\pmatrix{F(r) \cr G(r)}:
\ee
the functions $F, G$ satisfy coupled first order equations
\bb
EF&=&\hbar (1-\frac{r_h}{r})G'\nonumber\\
EG&=&-\hbar (1-\frac{r_h}{r})F',
\ee
which can be solved {\it exactly}. The na\"{\i}ve incoming solution is
\bb
F=-iG\propto -ie^{-i\frac{E}{\hbar}\int {dr\over 1-\frac{r_h}{r}}},
\ee
while the na\"{\i}ve outgoing solution is
\bb
F=iG\propto ie^{i\frac{E}{\hbar}\int {dr\over 1-\frac{r_h}{r}}}.
\ee
Both solutions contain damping factors $\exp (-\pi Er_h/\hbar)$
which lead to factors $\exp (-2\pi Er_h/\hbar)$ in the probability.
To make the absorption complete, as in the case of the scalar,
a factor of $\exp (-iC/\hbar)$ can be used, whereupon the usual 
emission probability appears. But there are {\it no
corrections} to the dependence on $E$ or to the Hawking temperature.
For more involved angular dependence, the equations for $F, G$ have
extra terms \cite{brill} and corrections to the above solution start from $S_2$,
with $S_n\propto E^{1-n}$ once again, changing the distribution in a
complicated $r$-dependent way. But one cannot interpret it as a
correction to the Hawking temperature.

To sum up, we have calculated corrections to the spherically
symmetric solution of the
Klein-Gordon equation in Schwarzschild and advanced Eddington-Finkelstein
coordinates and found that these corrections, which
alter the amplitude, do not involve corrections to
the Hawking temperature. For the Dirac equation, exact solutions can be
obtained for a simple angular dependence and in this case there
is no correction to the solution, while other solutions get altered,
but again without any change in the Hawking temperature.

\acknowledgments
We thank Amit Ghosh for discussions on this issue.

\end{document}